\documentclass[10pt]{article}
\usepackage[preprint]{tmlr}
\usepackage{amsmath,amssymb}
\usepackage{graphicx}
\usepackage{booktabs,tabularx,array}
\usepackage{url}
\usepackage{hyperref}
\hypersetup{colorlinks=true,linkcolor=blue,citecolor=blue,urlcolor=blue}
\makeatletter\g@addto@macro\UrlBreaks{\do\_\do\-\do\a\do\b\do\c\do\d\do\e\do\f\do\g\do\h\do\i\do\j\do\k\do\l\do\m\do\n\do\o\do\p\do\q\do\r\do\s\do\t\do\u\do\v\do\w\do\x\do\y\do\z}\makeatother
\DeclareUnicodeCharacter{00B9}{\textsuperscript{1}}
\DeclareUnicodeCharacter{010D}{\v{c}}
\providecommand{\tightlist}{\setlength{\itemsep}{0pt}\setlength{\parskip}{0pt}}
\title{Quantum kernels on breast data: concentration, surrogates, and two ways an advantage gets manufactured}
\author{\name Rana Saad \email ORCID 0009-0005-5806-2604\\ \addr Qanary Inc.
\AND
\name Achraf Toufah \email ORCID 0009-0006-6802-3905\\ \addr Qanary Inc.
\AND
\name Moulay Youssef El Hafidi \email myelhafidi@qanary.net\\ \addr Qanary Inc. (corresponding author; ORCID 0000-0001-7413-3293)}
\def\month{MM}\def\year{YYYY}\def\openreview{\url{https://openreview.net/forum?id=XXXX}}
\begin{document}
\maketitle
\begin{abstract}
Quantum-kernel advantage claims on breast data are usually a single number: one quantum arm, one comparator, one test split. We asked what that number measures once the full comparator set and selection history are visible. We report three active benchmarks, all exact classical statevector simulations of 8-qubit fidelity kernels with no quantum component in the inference path: a sample-efficiency sweep on Wisconsin Diagnostic Breast Cancer at ten training-set sizes; raw-versus-processed arms on 1,200 M-SYNTH simulated mammograms with a matched closed-form dequantisation surrogate, five classical methods at the same input and eleven kernel families; and a 30-seed selection-inflation measurement on BreastMNIST. A CMMD D1 clinical arm is withdrawn as a result arm because its frozen splits leak the label through patient serial order, a defect reported in a separate study \citep{r32}; it is retained only as a worked example. All three active arms are negative. On M-SYNTH raw an apparent +0.1736 (p = 5.2e-06) against a tuned RBF-SVM falls to +0.0019 (p = 0.937) against a random forest at the same input, and a Laplacian kernel wins on validation. On WDBC, with model selection confined to training data, the quantum kernel is below the selected classical arm at every training-set size, with 2 of 200 paired subsamples ahead. Best-of-30 seed selection adds +0.0210 on BreastMNIST. One result does not fit: at bandwidth 2pi the matched closed-form Gaussian surrogate failed to reproduce the M-SYNTH raw kernel (0.5410 against 0.7045, p = 6.3e-06). Comparator choice and selection each produce effects larger than any quantum-classical difference we measured; we give a reporting standard that removes both.
\end{abstract}

\section{Introduction}

A quantum-kernel result on clinical data, like the broader set of quantum-machine-learning claims on breast data audited in preparing this work \citep{r1,r2,r3,r4,r5}, is normally reported as a difference: the area under the receiver operating characteristic curve \citep{r6} of a fidelity-kernel support vector machine \citep{r7}, minus the AUC of a classical comparator, on a test split. The difference is small, the test split is small, and almost every degree of freedom that determines its sign lies outside the number: which classical method was chosen, how many configurations each arm received, whether the reported configuration was selected on the validation split or on the test split, and how many seeds were run and which one appears in the table. The generalisation results most often invoked in support of a sample-efficiency advantage for quantum learners \citep{r8,r9} were not derived for classical clinical data and do not establish one for it.

This paper measures those degrees of freedom on breast data, in our own code, and reports what is left of the difference when they are closed. We report three active benchmarks and one withdrawn clinical arm retained as a worked example. The active arms returned no advantage. The useful content is not the null results themselves; it is that the mechanisms that would convert them into positives are quantitatively larger than the quantity being measured, and that those mechanisms can be shown in our own artefacts.

The first mechanism is the comparator. On simulated raw mammograms, the best 8-qubit quantum kernel that could be selected on validation reached test AUC 0.7045 against a tuned RBF-kernel SVM at 0.5309, a difference of +0.1736 at p = 5.159e-06 (5.159074640941973e-06 at full precision). Taken alone this is a defensible-looking result, obtained on internally registered splits with an equal tuning budget and a paired DeLong test. A random forest on exactly the same eight dimensions reaches 0.7026, which leaves +0.0019 at p = 0.9369. The RBF-SVM was the second-weakest of the five classical methods run at that input, and its weakness was already visible on the validation split, where it also ranked fourth of five; the rankings by validation and by test AUC are identical for all five methods. No test-set information was needed to know that the comparator was weak.

The second mechanism is selection. On BreastMNIST, taking the maximum over 30 seeds of one fixed architecture rather than the mean inflates test AUC by 0.0210, which is the size of the entire advantage claimed by one of the published results audited in preparing this work. A previously reported CMMD test-set-selection artefact is no longer used as a result, because the CMMD D1 arm has been withdrawn after a serial-number gate showed that patient ordering predicts malignancy on the frozen split.

Both mechanisms produce effects an order of magnitude larger than the largest quantum-classical difference in our active data, and quantifying them is the contribution of this paper.

We position this work against the framework proposed by Lanes et al.~(arXiv:2506.20658) \citep{r10}, referred to throughout as the IBM/Pasqal framework; it is a proposed functional definition of quantum advantage, not a standard adopted by any standards body. Its criterion (ii) requires a separation stated relative to the best known classical methods and tested against ``both existing and emerging classical computational methods'' (Lanes et al., Section II.B, p.~2); its criterion (i) requires rigorously validated correctness. Quantum kernels sit in the framework's third algorithmic family, estimation of expectation values, where the framework cites the ZZ feature map \citep{r7} directly. The framework also supplies the sentence under which a result such as ours is a contribution rather than a complaint: should a classical approach later supersede a previously reported quantum advantage benchmark, the outcome ``should not be viewed as a failure; rather, it exemplifies constructive behavior that drives progress in both classical and quantum computation'' (Lanes et al., Section II.B, p.~2). We apply that standard deliberately, on clinical data, to our own results.

Two qualifications belong in the first section rather than the last. First, the framework was written for physics and optimisation workloads whose quality score is a deterministic functional of the output; a held-out AUC is a noisy estimate of a population quantity with no one-sided guarantee, and the framework offers no statistical-inference layer for that case. Supplying one, in the form of a minimum detectable effect compared against headroom, paired tests and pre-declared selection discounts, is the methodological contribution here, and it is ours rather than the framework's. Second, an 8-qubit kernel is roughly two orders of magnitude inside the classically simulable regime that the framework identifies for error-mitigated hardware (approximately 100 qubits with thousands of two-qubit gates) \citep{r10,r11,r12,r13,r14,r15}, so nothing in this paper could have demonstrated an advantage even had the sign gone the other way. What it can do is measure how a null becomes a positive.

Three statements hold throughout and are not repeated at every occurrence: every quantum run reported here is a classical statevector simulation on 8 qubits; no quantum hardware was used; and there is no quantum component in the inference path.

\section{Methods}

\subsection{Design common to all arms}

Three active arms, three datasets and one quantum model family were used: the fidelity kernel of an 8-qubit feature map \citep{r7}, evaluated by exact statevector simulation and used as a precomputed kernel in a support vector machine. The ZZ feature-map statevector is constructed in closed form, as a diagonal phase from the single-qubit terms and the pairwise products alternated with a Walsh-Hadamard transform for the Hadamard layer, and the Gram matrix is one matrix multiplication over the state array. There is no sampling, no shot noise, no device calibration and no error mitigation, because there is no device. Transpiled circuit depth and two-qubit gate count are not reported, because no circuit is transpiled in a simulation-only study; they belong to the consolidation run proposed in Section 4.3.

Of the three active arms, only the M-SYNTH arm (Section 2.3) was pre-registered. Throughout this paper ``pre-registered'' and ``registered'' refer to a dated internal protocol document frozen before the corresponding run and deposited with the released code; the registration was not lodged on a public registry. A second internal registration, written for the subsequently withdrawn clinical arm, is cited where its power calculations bear on the present results.

The encodings compared were the ZZ feature map with a nearest-neighbour entanglement graph, the ZZ feature map with an all-to-all graph, and angle encoding, which is a product state with no entanglement and whose kernel is a classical shift-invariant product of cosines. Bandwidths on the M-SYNTH arm were pi/8, pi/4, pi/2, pi and 2pi, with two repetitions throughout.

The matched dequantisation surrogate is the anisotropic Gaussian with metric M = I + pi\^{}2 Q, where Q is the signless Laplacian of the entanglement graph \citep{r16}. In the attribution audit it is fixed to the same edge set and the same bandwidth as the winning quantum configuration, and its regularisation constant alone is tuned on validation. This distinction matters and is the subject of Section 3.3.

\subsection{Arm A: WDBC sample efficiency}

The Wisconsin Diagnostic Breast Cancer dataset \citep{r17} comprises 569 cases with 30 features. A fixed held-out test set was used; all preprocessing (standardisation, PCA, min-max scaling) was fitted inside each training subsample only; 15-20 paired subsamples were drawn per point, with every arm receiving the identical subsample; and hyperparameters, including the quantum bandwidth and the choice among the classical models, were tuned by inner cross-validation on the training subsample for every arm, with the selected model scored once on the test set. Three regimes were run: all 30 features; a de-saturated feature block (standard-error columns only); and a rare positive class, with 3-35 malignant examples against 120 benign. The classical arms were logistic regression and RBF-SVM on PCA features, and RBF-SVM on all 30 features. This arm was not registered: it was run before the protocol that governs the follow-up experiment was written, and the fact is recorded as a deviation in the internal registration document.

\subsection{M-SYNTH, raw and processed}

M-SYNTH \citep{r18} (FDA Regulatory Science Tool RST24AI06.01, CC0-1.0), generated with the VICTRE in-silico pipeline \citep{r19}, was used with PresentationIntentType = FOR PROCESSING verified by direct DICOM header read. The cohort comprised 1,200 cases: 4 breast densities x 2 dose levels x 150, balanced 600 lesion-present / 600 lesion-absent; one 256x256 patch per phantom at the native 0.085 mm pitch, with control patches placed at coordinates drawn from the positive lesion-coordinate distribution of the same cell so that patch position cannot carry the label. The lesion is a spiculated mass of radius 5.0 mm and lesion density 1.06. Splits were stratified on density, dose and label: train 672, validation 176, test 352 (176 positive). The recorded checksums match the delivered split files byte for byte.

Both representations use the identical exposures and the identical patches. RAW is 16-bit detector-linear counts as delivered. PROC applies a representative presentation chain, comprising logarithmic attenuation, windowing to the 1st-99.5th percentile and requantisation to 8 bits, which is information-reducing only and includes no sharpening. Features are reduced to 8 dimensions by PCA fitted on the training split only, retaining 0.9614 of variance on raw.

Three passes were run, and their order is part of the result.

\begin{enumerate}
\def\labelenumi{\arabic{enumi}.}
\tightlist
\item
  First pass: the best quantum configuration against the best RBF-SVM configuration, both selected on validation, with 60 configurations per arm, plus 20 paired bootstrap resamples and the raw-versus-processed interaction test. The pre-specified stopping rule was fixed in code at a minimum detectable delta-AUC of 0.0406.
\item
  Attribution audit \citep{r20}: the matched surrogate, five classical methods at the same 8-dimensional input (RBF-SVM, histogram gradient boosting, logistic regression, random forest, k-nearest neighbours), the best classical model on the full feature set, and Gram-concentration diagnostics.
\item
  Kernel-family sweep: eleven kernel families at the same input, added after the audit and not registered; the addition is recorded as a deviation in the internal registration document.
\end{enumerate}

\subsection{BreastMNIST selection inflation}

The MedMNIST v2 official split \citep{r21} was used: 546 train / 78 validation / 156 test (42 positive, 114 negative). Five classical model families were run on raw pixels and on HOG features, with every hyperparameter tuned on validation only, the test split touched once, and 4,000-sample bootstrap confidence intervals. Separately, one fixed architecture was run over 30 seeds and the full distribution recorded. There was no registration; this is a measurement of a reporting artefact, not a quantum-versus-classical test.

\subsection{Statistics}

Paired DeLong tests \citep{r22} with a midrank implementation are reported as delta-AUC with its standard error and two-sided p-value. In the WDBC sweep the subsamples share one test set and overlap in their training data, so they are not treated as independent replicates: the mean paired difference across subsamples is given a two-level bootstrap interval (2,000 replicates; subsamples and test cases resampled together, test cases stratified by label), and a Nadeau-Bengio corrected resampled t-test is reported as a conservative sensitivity analysis. Uncorrected Wilcoxon tests are not used for inference. Other 95\% confidence intervals are obtained by bootstrap. Minimum detectable effects were taken from the power tables in the two internal registrations.

The analyses are ranked. The registered M-SYNTH first-pass comparisons (raw and processed) are confirmatory and are Holm-adjusted as a pair \citep{r23}. The WDBC sweep is exploratory and is Holm-adjusted within each regime. Everything else (the M-SYNTH attribution audit, the kernel-family sweep and the BreastMNIST measurements) is exploratory, and its p-values are reported unadjusted as descriptive quantities. The selection artefacts are descriptive quantities and are not tested.

Concentration diagnostics are computed on the training Gram matrix. Each entry is the fidelity kernel, the squared overlap of two feature-map states; the states are normalised, so the diagonal is 1, and no further normalisation is applied. The mean off-diagonal entry is the average over the n(n - 1) ordered pairs of distinct training cases. The Haar floor is 1/2\^{}n for n qubits, the expected fidelity between two independent Haar-random states (1/2\^{}8 = 0.00390625 here), and the ratio of the mean off-diagonal entry to that floor measures how far the kernel sits above it. The effective rank is the exponential of the Shannon entropy of the normalised eigenvalue spectrum of the training Gram matrix, with eigenvalues divided by their sum and negative rounding errors set to zero; it is reported with its fraction of the training-set size.

\section{Results}

\subsection{Arm A: the kernel concentrates, and the deficit shrinks as N grows}

Table 1 gives the WDBC sample-efficiency results, and Figure 2 shows the corresponding curves for the three regimes.

\begin{figure}[!tbp]\centering
\includegraphics[width=\textwidth]{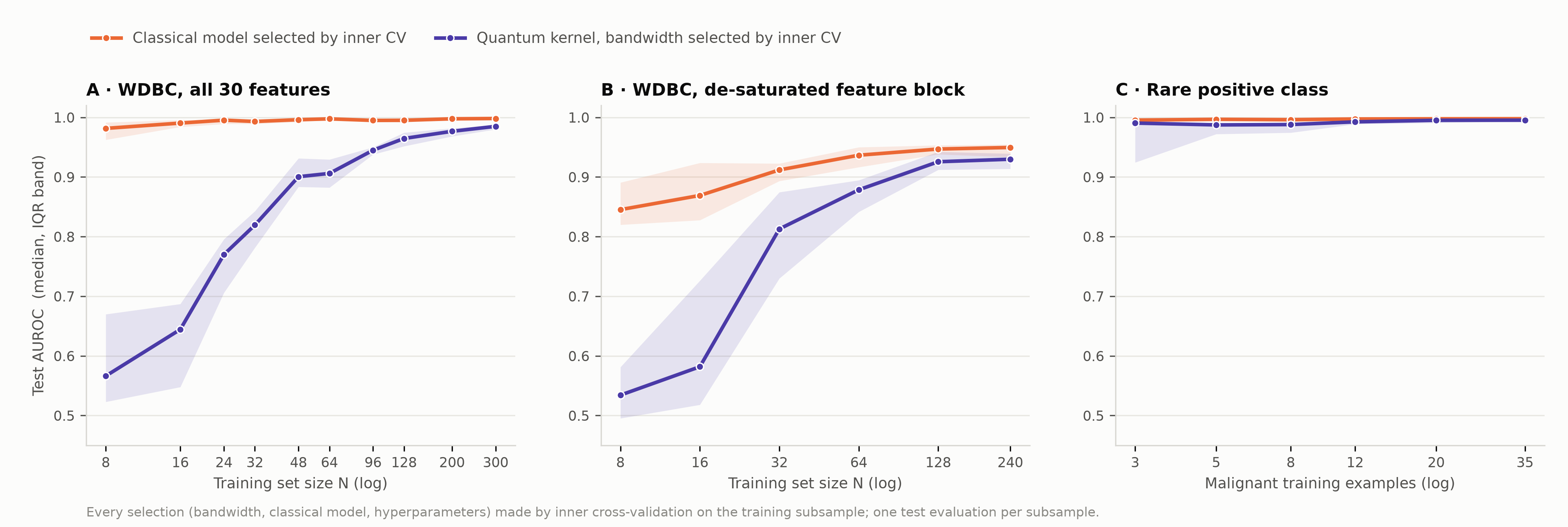}
\caption{WDBC sample-efficiency curves in three regimes: all 30 features; a de-saturated feature block (standard-error columns only); and a rare positive class. Test AUC against training-set size N for the quantum kernel, with its bandwidth chosen by inner cross-validation, and for the classical model chosen by inner cross-validation: median and interquartile range over 20 (all features) or 15 (other regimes) paired subsamples. The quantum median is below the classical median at every N in all three regimes; Table 1 gives the paired differences with two-level bootstrap intervals for the all-feature regime.}
\label{fig:2}
\end{figure}

\begin{table}[!tbp]
\centering\small
\caption{WDBC sample efficiency: quantum kernel against the best classical arm at ten training-set sizes. Twenty paired subsamples per size. In each subsample the quantum bandwidth (1.0, 0.2 or 0.05) and the classical model (logistic regression on PCA features, or RBF-SVM on PCA features or on all 30 features) were chosen by inner cross-validation on the training subsample, and the chosen model was scored once on the held-out test set. Medians with interquartile ranges; mean paired difference (quantum minus classical) with a two-level bootstrap 95\% interval that resamples subsamples and test cases together; bootstrap p-values Holm-adjusted across the ten sizes, where 0.005 is the resolution limit of 2,000 replicates after adjustment. Exploratory: this arm was not registered.}
\label{tab:1}
\begin{tabularx}{\textwidth}{l>{\raggedright\arraybackslash}X>{\raggedright\arraybackslash}X>{\raggedright\arraybackslash}X>{\raggedright\arraybackslash}Xl}
\toprule
\textbf{N} & \textbf{Quantum median {[}IQR{]}} & \textbf{Classical median {[}IQR{]}} & \textbf{Mean paired Delta {[}95\% CI{]}} & \textbf{Subsamples quantum ahead} & \textbf{p (Holm)} \\
\midrule
8 & 0.5662 {[}0.5229, 0.6698{]} & 0.9815 {[}0.9624, 0.9915{]} & -0.3940 {[}-0.4439, -0.3415{]} & 0 / 20 & 0.005 \\
16 & 0.6443 {[}0.5478, 0.6871{]} & 0.9904 {[}0.9840, 0.9956{]} & -0.3628 {[}-0.4112, -0.3148{]} & 0 / 20 & 0.005 \\
24 & 0.7699 {[}0.7059, 0.7957{]} & 0.9953 {[}0.9889, 0.9965{]} & -0.2493 {[}-0.2983, -0.2014{]} & 0 / 20 & 0.005 \\
32 & 0.8194 {[}0.7815, 0.8427{]} & 0.9931 {[}0.9890, 0.9968{]} & -0.1812 {[}-0.2228, -0.1460{]} & 0 / 20 & 0.005 \\
48 & 0.9004 {[}0.8835, 0.9313{]} & 0.9961 {[}0.9922, 0.9980{]} & -0.1027 {[}-0.1399, -0.0713{]} & 0 / 20 & 0.005 \\
64 & 0.9060 {[}0.8822, 0.9293{]} & 0.9974 {[}0.9958, 0.9985{]} & -0.1001 {[}-0.1377, -0.0669{]} & 0 / 20 & 0.005 \\
96 & 0.9449 {[}0.9378, 0.9499{]} & 0.9948 {[}0.9930, 0.9978{]} & -0.0474 {[}-0.0710, -0.0271{]} & 1 / 20 & 0.005 \\
128 & 0.9645 {[}0.9516, 0.9743{]} & 0.9950 {[}0.9918, 0.9982{]} & -0.0306 {[}-0.0492, -0.0154{]} & 1 / 20 & 0.005 \\
200 & 0.9766 {[}0.9681, 0.9818{]} & 0.9975 {[}0.9955, 0.9982{]} & -0.0220 {[}-0.0381, -0.0099{]} & 0 / 20 & 0.005 \\
300 & 0.9849 {[}0.9799, 0.9874{]} & 0.9980 {[}0.9977, 0.9988{]} & -0.0143 {[}-0.0274, -0.0049{]} & 0 / 20 & 0.005 \\
\bottomrule
\end{tabularx}
\end{table}

With every selection confined to the training data, the quantum arm was ahead of the selected classical arm in 2 of 200 paired subsamples, and the mean paired difference was negative at every size, with two-level bootstrap intervals that exclude zero at all ten sizes. The deficit narrows as N grows, from -0.3940 at N = 8 to -0.0143 at N = 300; this is the opposite shape to a sample-efficiency advantage, which would require the quantum arm to lead at small N and converge at large N (Figure 2). In the de-saturated and rare-class regimes the quantum median is also below the classical median at every size; the intervals exclude zero at 5 of 6 sizes and at 3 of 6 sizes respectively (Holm-adjusted p 0.003 to 0.088 and 0.006 to 0.314), and the quantum arm was ahead in 12 and 14 of 90 paired subsamples.

An earlier version of Table 1 took, in each subsample, the larger of the test AUCs across bandwidths and across classical models. That reading selected on the test set and is withdrawn; it gave the same direction at every size (mean paired differences -0.3172 at N = 8 to -0.0147 at N = 300). The subsamples are not independent replicates: they share one test set and overlap in their training data. The uncorrected Wilcoxon p-values of the earlier version therefore overstated the evidence, and a Nadeau-Bengio corrected resampled t-test, a deliberately conservative alternative, is not significant at any size after Holm adjustment (smallest adjusted p 0.083, at N = 300). The result of this arm is the direction and size of the deficit, not its p-value.

The headroom rule disqualifies this arm as a test of advantage, as distinct from a description of one: the classical median is already 0.9815 at N = 8. The internal registration written two months later for the clinical arm states the consequence for WDBC explicitly: tuned classical models reach 0.994 AUC, leaving 0.005 of headroom against a smallest-detectable difference of 0.014, so that no advantage is detectable there even in principle. We report arm A as a mechanism result only.

Table 2 shows the mechanism. The quantum arm's bandwidth is what moves it, and the direction in which the tuner drives it is diagnostic of exponential kernel concentration \citep{r24}.

\begin{table}[!tbp]
\centering\small
\caption{WDBC: Gram-matrix concentration against the Haar floor, and the AUROC it buys. Gram diagnostics are computed on the training Gram matrix at one configuration, N = 300 and seed 0 (298 cases, because the training pool holds 148 malignant cases); definitions are given in Section 2.5. The AUROC column is the median over 20 subsamples at N = 300.}
\label{tab:2}
\begin{tabularx}{\textwidth}{>{\raggedright\arraybackslash}X>{\raggedright\arraybackslash}X>{\raggedright\arraybackslash}X>{\raggedright\arraybackslash}X>{\raggedright\arraybackslash}X}
\toprule
\textbf{Bandwidth} & \textbf{Mean off-diagonal Gram} & \textbf{Ratio to the 1/2\^{}8 floor (0.00390625)} & \textbf{Effective rank (fraction of 298)} & \textbf{Median test AUROC at N = 300} \\
\midrule
1.0 (default scale) & 0.0060 & 1.52 & 295.2 (0.991) & 0.8172 \\
0.2 & 0.0468 & 11.98 & 207.3 (0.696) & 0.9652 \\
0.05 & 0.6434 & 164.71 & 6.5 (0.022) & 0.9849 \\
\bottomrule
\end{tabularx}
\end{table}

At the default encoding scale the mean off-diagonal Gram entry sits a factor 1.52 above the 1/2\^{}8 Haar floor: the kernel matrix is effectively the identity, every training point appears equidistant from every other, and a model fitted on it memorises the training set. Bandwidth tuning recovers most of the performance, and it does so by driving the kernel away from the quantum regime: the inner-cross-validation tuner selects the smallest bandwidth in the search range in most subsamples (15 to 20 of 20 at every N of 24 or more in the all-feature regime; 10 to 13 of 15 at N of 32 or more in the de-saturated regime, and 11 to 15 of 15 in the rare-class regime), which is the limit in which the ZZ feature map linearises toward a trivial classical kernel. The more classical the encoding is made, the better it performs.

An internal summary quoted the default-scale mean off-diagonal entry as 0.005. That figure is consistent with the released value, 0.005952528989071272, truncated to three decimal places, and no separate measurement giving 0.005 exists in the released files; re-running the released code at the same configuration reproduces 0.005952528989071272 to all digits. Table 2 gives it rounded, 0.0060.

\subsection{Mechanism one: the comparator decides the result}

Table 3 and Figure 1a show the same quantum kernel against different classical comparators on the same frozen test split.

\begin{figure}[!tbp]\centering
\includegraphics[width=\textwidth]{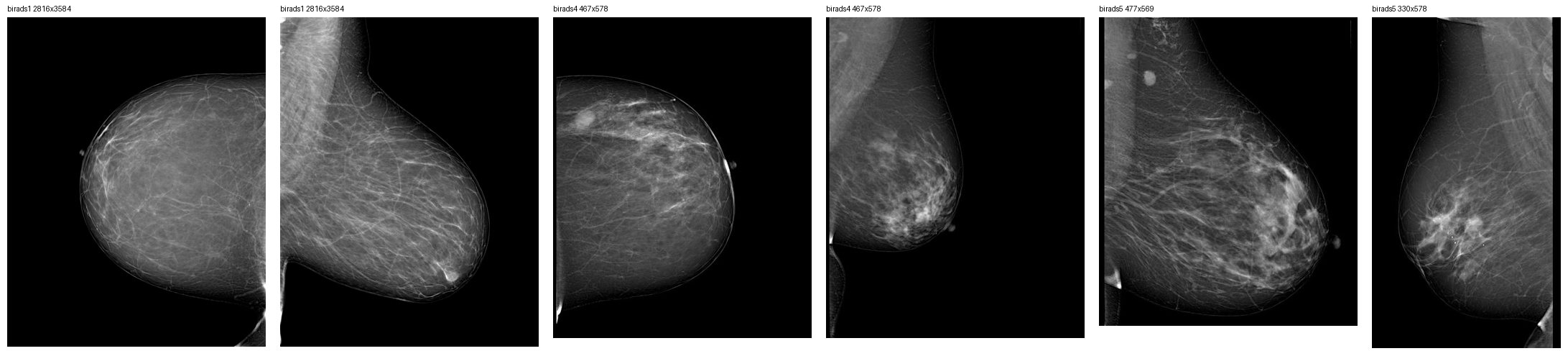}
\caption{Two mechanisms by which a quantum-kernel advantage can be manufactured. (a) M-SYNTH raw arm, frozen test split (n = 352), all arms at the same 8-dimensional input: the quantum kernel (test AUC 0.7045) is +0.1736 above the registered RBF-SVM comparator (p = 5.2e-06) and +0.0019 above a random forest at the identical input (p = 0.94). Stems start at AUC 0.50. (b) BreastMNIST, 30 seeds of one fixed classical architecture on the official 156-image test split: the distribution of test AUC has mean 0.7521 and maximum 0.7732, so best-of-30 reporting adds +0.0210. All quantum values are exact classical statevector simulations of an 8-qubit circuit.}
\label{fig:1}
\end{figure}

\begin{table}[!tbp]
\centering\small
\caption{M-SYNTH: the same quantum kernel against different classical comparators, same frozen test split (n = 352), same 8-dimensional input. Validation AUC is shown because it is what a registered protocol is allowed to select on.}
\label{tab:3}
\begin{tabularx}{\textwidth}{>{\raggedright\arraybackslash}Xll>{\raggedright\arraybackslash}Xl}
\toprule
\textbf{Arm at the same 8-dimensional input} & \textbf{Validation AUC} & \textbf{Test AUC} & \textbf{Quantum minus this arm} & \textbf{p} \\
\midrule
Quantum, ZZ all-to-all, beta = 2pi & 0.6872 & 0.7045 & --- & --- \\
Random forest & 0.7465 & 0.7026 & +0.0019 & 0.9369 \\
Gradient boosting (HistGB) & 0.7212 & 0.6542 & --- & --- \\
k-nearest neighbours & 0.6176 & 0.5755 & --- & --- \\
Matched closed-form surrogate & --- & 0.5410 & +0.1635 & 6.349e-06 \\
SVM, RBF kernel & 0.5718 & 0.5309 & +0.1736 & 5.159e-06 \\
Logistic regression & 0.4616 & 0.5044 & --- & --- \\
Best classical on the FULL feature set & --- & 0.7551 & -0.0506 & 0.1052 \\
\bottomrule
\end{tabularx}
\end{table}

Reading the first and last rows against each other is the whole of mechanism one. Against the registered RBF comparator the quantum kernel is ahead by +0.1736 at p = 5.159e-06, 4.3 times the pre-declared minimum detectable effect of 0.0406, reproduced in 20 of 20 paired resamples. Against a random forest at the identical input it is ahead by +0.0019 at p = 0.9369. Against the best classical model on the full feature set it is behind by 0.0506 (p = 0.1052).

The ranking of the five same-input classical methods is identical on validation and on test: random forest \textgreater{} gradient boosting \textgreater{} kNN \textgreater{} RBF-SVM \textgreater{} logistic regression on both. The comparator that manufactured the effect was therefore identifiable as the second-weakest of the five without consulting the test split. A protocol that fixes its comparator by name rather than as the best of a declared set selected on validation will reproduce this failure every time, and registration alone does not prevent it.

\begin{table}[!tbp]
\centering\small
\caption{M-SYNTH raw: eleven kernel families at the same input, ranked by test AUC. Configuration counts are per family and are not equal across families.}
\label{tab:4}
\begin{tabularx}{\textwidth}{>{\raggedright\arraybackslash}Xlll}
\toprule
\textbf{Kernel} & \textbf{Configurations} & \textbf{Validation AUC} & \textbf{Test AUC} \\
\midrule
QSVM, ZZ all-to-all & 20 & 0.6872 & 0.7045 \\
Laplacian & 60 & \textbf{0.7350} & 0.7009 \\
Laplacian-ARD & 60 & 0.7104 & 0.6768 \\
QSVM, ZZ nearest-neighbour & 20 & 0.6599 & 0.5950 \\
Polynomial d = 4 & 8 & 0.5440 & 0.5387 \\
Polynomial d = 3 & 8 & 0.5413 & 0.5350 \\
QSVM, angle (no entanglement) & 20 & 0.5708 & 0.5332 \\
RBF (isotropic) & 60 & 0.5718 & 0.5309 \\
Polynomial d = 2 & 8 & 0.5315 & 0.5057 \\
Linear & 4 & 0.4623 & 0.5039 \\
RBF-ARD (per-feature weights) & 60 & 0.5584 & 0.5027 \\
\bottomrule
\end{tabularx}
\end{table}

A non-Gaussian kernel is the second mandatory comparator that the published literature usually omits, and on this data it is decisive (Table 4). The Laplacian kernel reaches 0.7009 on test, within 0.0036 of the quantum kernel, and it wins on validation, 0.7350 against 0.6872. A protocol that selects its reported arm on validation, as both of our registrations require, would have reported the Laplacian kernel as the winner and the quantum kernel as the runner-up. The quantum arm's test-set lead over the Laplacian kernel exists only because the validation-selected ordering reverses on the test split, which is exactly the kind of difference a 352-case test set cannot resolve.

\begin{table}[!tbp]
\centering\small
\caption{M-SYNTH: the pre-specified stopping rule, before and after the audit. Threshold: delta-AUC \textgreater{} 0.0406 with p \textless{} 0.05. Holm-adjusted over the two registered first-pass comparisons, the first-pass p-values are 1.032e-05 (raw) and 0.0211 (processed); no verdict changes.}
\label{tab:5}
\begin{tabularx}{\textwidth}{l>{\raggedright\arraybackslash}Xlll}
\toprule
\textbf{Representation} & \textbf{Comparator} & \textbf{delta-AUC} & \textbf{p} & \textbf{Stopping rule} \\
\midrule
Raw & SVM-RBF only (first pass) & +0.1736 & 5.159e-06 & would PASS \\
Raw & best same-input classical (random forest) & +0.0019 & 0.9369 & \textbf{FAIL} \\
Raw & best classical on full features & -0.0506 & 0.1052 & \textbf{FAIL} \\
Processed & SVM-RBF only (first pass) & +0.1023 & 0.0211 & would PASS \\
Processed & best same-input classical (kNN) & +0.0211 & 0.5992 & \textbf{FAIL} \\
Processed & best classical on full features & -0.0121 & 0.7574 & \textbf{FAIL} \\
\bottomrule
\end{tabularx}
\end{table}

The registered question, whether the quantum-minus-classical gap differs between raw and processed inputs, is answered in the negative (Table 5): the interaction is +0.0713 at p = 0.2226 against the first-pass comparator, and against the corrected baselines the two gaps are +0.0019 and +0.0211, neither significant. Raw data does not rescue quantum kernels.

\subsection{One result that does not fit: the matched surrogate failed at beta = 2pi}

The closed-form dequantisation surrogate is a second-order approximation: it replaces the ZZ fidelity kernel with an anisotropic Gaussian in coordinate differences, so it captures pairwise phase structure and assumes the kernel is a smooth, stationary function of those differences. On M-SYNTH raw, at the largest bandwidth in the grid, that assumption did not hold (Table 6, Figure 3).

\begin{figure}[!tbp]\centering
\includegraphics[width=\textwidth]{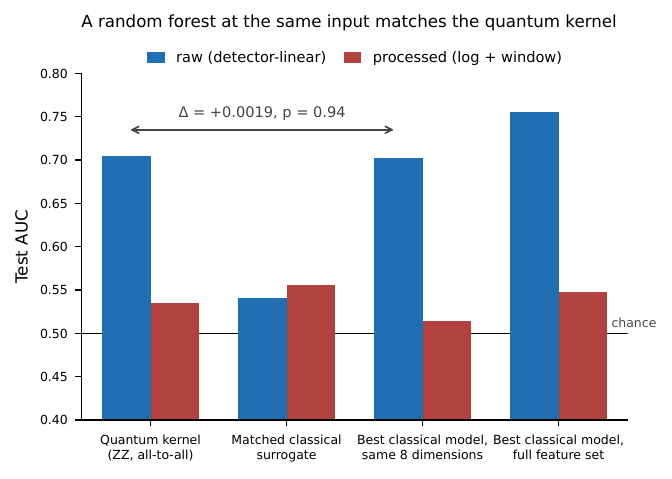}
\caption{M-SYNTH raw arm: the quantum kernel (test AUC 0.7045) against its matched closed-form surrogate at the same feature map and bandwidth (0.5410), a random forest at the same eight dimensions (0.7026) and the best classical model on the full feature set (0.7551), shown for the raw and processed representations. The matched surrogate did not reproduce the kernel at beta = 2pi, whereas the random forest matches it to within 0.0019.}
\label{fig:3}
\end{figure}

\begin{table}[!tbp]
\centering\small
\caption{M-SYNTH: the matched surrogate and the concentration diagnostics for the winning quantum configuration (ZZ all-to-all, beta = 2pi, 8 qubits).}
\label{tab:6}
\begin{tabularx}{\textwidth}{>{\raggedright\arraybackslash}X>{\raggedright\arraybackslash}X>{\raggedright\arraybackslash}X}
\toprule
\textbf{Quantity} & \textbf{Raw} & \textbf{Processed} \\
\midrule
Quantum kernel, test AUC & 0.7045 & 0.5355 \\
Matched surrogate, same map and bandwidth, test AUC & \textbf{0.5410} & 0.5557 \\
Quantum minus matched surrogate & +0.1635 (p = 6.349e-06) & -0.0202 (p = 0.5792) \\
Mean off-diagonal Gram & 0.0293 & 0.0244 \\
1/2\^{}8 Haar floor & 0.0039 (exactly 0.00390625) & 0.0039 \\
Ratio to the floor & 7.51 & 6.25 \\
Effective-rank fraction of the training Gram & 0.2997 & 0.5296 \\
PCA variance retained at 8 dimensions & 0.9614 & --- \\
\bottomrule
\end{tabularx}
\end{table}

On raw data the matched surrogate reached 0.5410 against the kernel's 0.7045, p = 6.349e-06: the closed-form classical approximation did not reproduce the kernel. Concentration was moderate rather than severe, with the mean off-diagonal entry a factor 7.51 above the Haar floor and the effective rank 0.2997 of the training set, that is, about 201 of 672 directions, which is why an all-to-all map at high bandwidth was usable at all here.

Three points must be made about this result, because it is the one number in the paper that points the other way.

First, it does not rehabilitate the quantum arm: a random forest at the same input already matches the kernel to +0.0019. Hardness and usefulness are independent axes, and a kernel that resists a closed-form approximation while a tree ensemble reproduces its accuracy has demonstrated the former and not the latter.

Second, it bounds a claim made elsewhere more narrowly than it was made. The anisotropic Gaussian is a good approximation at small bandwidth and evidently not at beta = 2pi. The statement that the ZZ kernel is its own classical surrogate \citep{r16} is supported in the small-bandwidth regime and is not established at the top of the bandwidth grid. This is a statement about one closed-form family, the anisotropic Gaussian with the signless-Laplacian metric; it is not a limit on dequantisation in general, because no learned classical kernel was tested.

Third, the test most likely to overturn it lies outside the present study. A learned classical kernel, such as random Fourier features with learned frequencies or a neural kernel, fitted to the quantum Gram matrix is listed in the internal registration and is left to future work. If a learned classical kernel reproduces the Gram matrix to within the minimum detectable effect, the kernel is dequantised in practice even though no closed form reproduces it. It is the obvious objection to this result, and we name it here.

Two analyses report a surrogate AUC for the raw arm, and both values are retained: 0.5520402892561983 with p = 2.050205488795926e-05 from the first-pass analysis, and 0.5409510588842975 with p = 6.3486429348046396e-06 from the attribution audit. These are different quantities rather than inconsistent measurements. The first-pass script selects the best surrogate configuration on validation across the whole surrogate pool, whereas the audit script fixes the surrogate to the winning quantum configuration's map and bandwidth and tunes only its regularisation. The matched value, 0.5410, is the one that answers the dequantisation question and is the one used above; the selected value is reported here so that the two are not confused.

\subsection{Mechanism two: selection, measured on BreastMNIST}

On BreastMNIST, 30 seeds of one fixed architecture give a mean test AUC of 0.7521 with SD 0.0108, a maximum of 0.7732, and an inflation of 0.0210 (Figure 1b). On a 156-image test set, reporting the best of 30 seeds rather than the mean adds +0.0210, which is the size of the entire advantage claimed by one of the published results audited in preparing this work \citep{r5}, whose reported gap over its own classical convolutional network was +0.023.

\begin{table}[!tbp]
\centering\small
\caption{BreastMNIST: ten tuned classical arms on the official 156-image test split, with bootstrap 95\% confidence intervals.}
\label{tab:7}
\begin{tabularx}{\textwidth}{>{\raggedright\arraybackslash}Xllll}
\toprule
\textbf{Model} & \textbf{Features} & \textbf{Validation AUC} & \textbf{Test AUC} & \textbf{95\% CI} \\
\midrule
HistGradientBoosting & pixels & 0.9215 & 0.8615 & {[}0.7822, 0.9261{]} \\
SVM-RBF & pixels & 0.9449 & 0.8546 & {[}0.7851, 0.9125{]} \\
MLP & pixels & 0.8855 & 0.8494 & {[}0.7782, 0.9103{]} \\
RandomForest & pixels & 0.8647 & 0.8428 & {[}0.7672, 0.9071{]} \\
SVM-RBF & HOG & 0.9215 & 0.8347 & {[}0.7514, 0.9022{]} \\
RandomForest & HOG & 0.8881 & 0.7978 & {[}0.7116, 0.8706{]} \\
LogisticRegression & pixels & 0.8571 & 0.7943 & {[}0.7118, 0.8668{]} \\
HistGradientBoosting & HOG & 0.9006 & 0.7746 & {[}0.6785, 0.8550{]} \\
MLP & HOG & 0.8429 & 0.7533 & {[}0.6517, 0.8472{]} \\
LogisticRegression & HOG & 0.8287 & 0.7525 & {[}0.6469, 0.8444{]} \\
\bottomrule
\end{tabularx}
\end{table}

Table 7 and Figure 4 must be read in both directions. The best tuned classical model here, at 0.8615, is below the published convolutional and quantum convolutional results on this benchmark, and no matched convolutional network was trained, so this is not a demonstration that a tuned classical model equals a quantum one. What it establishes is the scale of the uncertainty. The 95\% interval of an off-the-shelf gradient-boosting model on raw pixels spans 0.144 of AUC and contains the best reported quantum value; every interval in Figure 4 overlaps every other. At 156 test images this benchmark cannot separate a quantum convolutional kernel from a tuned classical baseline, and the seed-selection measurement shows that it can be made to appear to. Because no matched convolutional network was trained, the arm does not compare our classical models with the published quantum results on equal terms, and we draw no conclusion of classical or quantum superiority from it.

\begin{figure}[!tbp]\centering
\includegraphics[width=\textwidth]{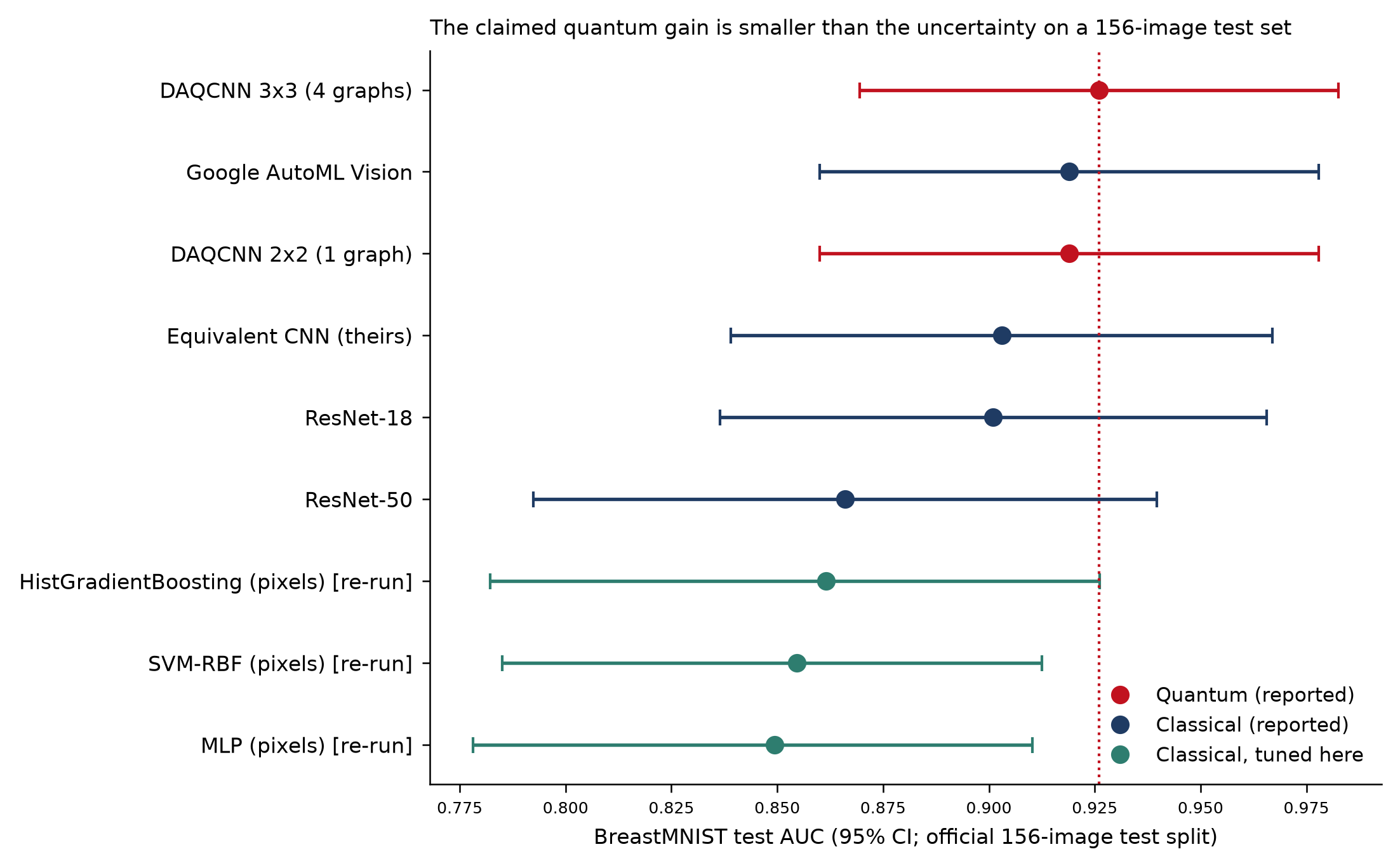}
\caption{BreastMNIST test AUC with 95\% confidence intervals for the reported quantum arms, the reported classical arms and the re-run tuned classical arms of Table 7 on the official 156-image test split. All intervals overlap.}
\label{fig:4}
\end{figure}

\section{Discussion}

\subsection{What the three active nulls jointly establish}

The active arms do not establish that quantum kernels cannot work. Three active benchmarks on breast data, at 8 qubits, on features of 8 to 30 dimensions, returned no advantage. The withdrawn CMMD arm is not counted as a benchmark result because its frozen D1 splits leak the label through patient serial order. The active-arm result is a statement about this model family at this scale on these representations. It is not a statement about quantum machine learning in general, and we have no basis for one.

What the active arms establish jointly is sharper and more portable: the decisions that are not in the headline number are larger than the number. Comparator choice moved an effect by 0.1717 of AUC (from +0.1736 against SVM-RBF to +0.0019 against a random forest, with the same quantum arm, the same input and the same test split). Best-of-30 seed selection moved it by 0.0210 on BreastMNIST. The withdrawn CMMD arm supplies a separate warning: even an internally registered clinical benchmark can be undone by a dataset-ordering defect. The largest quantum-classical difference anywhere in the active data is 0.0211 and is not significant. Any protocol that leaves these three decisions undeclared is not measuring a quantum effect; it is measuring its own degrees of freedom.

\subsection{CMMD as a worked example of a dataset-ordering defect}

CMMD D1, obtained through The Cancer Imaging Archive \citep{r25}, is no longer a result arm in this paper. It is a worked example of the paper's central warning: apparent performance can be manufactured by a dataset defect even after a careful registration has caught an earlier defect. The first defect was known before training: the D2 sub-collection is 98.4\% malignant and a batch-prefix gate reached AUC 0.7753 (also reported in \citep{r32}), which motivated the D1 restriction. The second defect was found after the data had been assembled for analysis: inside D1 itself, the patient serial number alone predicts malignancy almost perfectly on the frozen split. That finding, with its values and its per-decile structure, is reported and characterised in a separate study of label leakage in public mammography datasets \citep{r32}, and is not re-reported here as a result of this paper. Because the original CMMD analysis code and feature table are not archived, rebuilding the arm would be a new study rather than a reproducible rerun. Consequently, CMMD AUCs, deltas, p-values and sensitivity figures are not reported as results anywhere in this paper.

\subsection{The reporting standard}

We propose one standard with four components, each of which independently erased an effect in our own measurements.

\begin{enumerate}
\def\labelenumi{\arabic{enumi}.}
\tightlist
\item
  Same-input best classical comparator, where the set is mandatory and the member is selected on validation. The set must contain a tuned tree ensemble (a random forest matched our kernel to +0.0019) and a tuned non-Gaussian kernel (the Laplacian kernel beat it on validation, 0.7350 to 0.6872). Nominating a comparator by name is the defect; the best of a declared set is the remedy. Every member, including the losers, should be reported.
\item
  Matched surrogate, not selected surrogate. The dequantisation control should be fixed to the winning quantum configuration's feature map and bandwidth. Our two surrogate analyses differ precisely on this point, and the matched version is the one that found the exception in Section 3.3.
\item
  Concentration against the 1/2\^{}n floor, with effective rank. The mean off-diagonal Gram entry, the floor, the ratio and the entropy-based effective rank should be reported for every configuration. The floor falls exponentially in n while a clinical dataset does not grow, so the qubit count must be justified as a floor-versus-signal calculation rather than as a capability claim.
\item
  A selection discount reported next to every headline. Two numbers are required: best-on-test minus best-on-validation for both arms, and the best-of-N inflation over at least 20 seeds. An effect smaller than its own selection discount is reported as no difference.
\end{enumerate}

Consolidating these into a single reporting standard, with the harness, seeds and split checksums deposited under a DOI, is the next step this paper proposes; a registration for that consolidation run will be frozen before the run is executed.

\subsection{Position under the IBM/Pasqal framework}

The framework proposed by Lanes et al.~(arXiv:2506.20658) \citep{r10} is a proposed functional definition, not an adopted standard, and we use it as a checklist rather than as an authority. Three of its elements bear directly on this work.

Its criterion (ii) requires that a separation be stated relative to the best known classical methods and tested against both existing and emerging ones. Mechanism one is a measurement of what happens when that requirement is relaxed by exactly one comparator. For classical tabular inputs, moreover, loading N records costs Theta(N) regardless of circuit \citep{r26}, so only an accuracy separation was available to us in the first place.

Its criterion (i) requires rigorously validated correctness. The framework's own authors apply this standard to the field's most prominent advantage claims and classical-simulation results \citep{r15,r27,r28,r29,r30} and conclude that several ``remain unsubstantiated under this criterion''. If the most-cited sampling demonstrations fail criterion (i), a breast-cancer quantum-kernel claim resting on a noiseless simulator with no error analysis fails it by a wider margin, including ours, which is why this paper claims no advantage of any kind.

Its Section II.B supplies the sentence that makes a result such as this a contribution: should a classical approach later supersede a previously reported quantum advantage benchmark, the outcome ``should not be viewed as a failure; rather, it exemplifies constructive behavior that drives progress in both classical and quantum computation.'' We quote it verbatim because it is the authors' own standard. We note that the same authors also forecast credible evidence of quantum advantage emerging within the next two years; that is a forecast in the authors' voice, from two institutions that sell quantum computing, which we repeat here only with that attribution attached and do not endorse.

Where the framework does not reach is statistical inference for learning claims. A variational energy is a one-sided bound, so a lower number is unambiguously better, and benchmarks such as the V-score \citep{r31} can be built on that property; a test-set AUC is a noisy estimate of a population quantity, and a lower-scoring model can be the better model. The framework offers no remedy for that case. The apparatus in this paper, namely a minimum detectable effect compared against headroom, paired DeLong tests, bootstrap intervals, at least 20 seeds and a pre-declared selection discount, is our substitute, and it is the part of this work we would defend most firmly.

\subsection{What this work does not license}

It does not license any claim that the product performs quantum computation; it does not, and no result here would support such a claim. It does not license a sample-efficiency claim in either direction. It does not license treating the beta = 2pi surrogate failure as evidence of a quantum effect, because a random forest already matches that kernel's accuracy, nor as a limit on dequantisation beyond the Gaussian family tested. And it does not license citing any earlier Qanary quantum-kernel or quantum-neural-network manuscript as evidence of quantum benefit; none of them is such evidence, and this paper supersedes their framing.

\section{Limitations}

\begin{enumerate}
\def\labelenumi{\arabic{enumi}.}
\tightlist
\item
  Two of the three active arms were not registered. The WDBC pilot ran before the protocol that governs its follow-up was written, and the BreastMNIST measurement had no registration at all. Under this paper's own rubric both score zero on registration. M-SYNTH is the only active result arm that was pre-registered, in the internal sense defined in Section 2.1.
\item
  The CMMD arm was withdrawn. Its results and splits are retained as evidence, but the original analysis code and feature table are not archived, and the frozen D1 splits leak the label through patient serial order. CMMD is therefore a worked example, not a result arm.
\item
  The most load-bearing comparator was post hoc. The Laplacian-kernel sweep was added after the audit and was not registered. Our strongest argument against our own quantum arm is a post-hoc analysis, and we label it as one.
\item
  Multiplicity correction was declared in both registrations and was not applied at the time. It is applied here after the fact: Holm across the two registered M-SYNTH first-pass comparisons and within each WDBC regime. All other p-values are unadjusted and exploratory.
\item
  Two registered audit components lie outside the scope of the present study: the matched-specificity check that no gain is a classical threshold shift, and the learned-surrogate test. Both are left to future work.
\item
  No hardware, no shots and no error bars on kernel entries. Every quantum run is an exact statevector simulation; entry-level uncertainty was never propagated into the Gram matrix, the decision function or the AUC, because there was no sampling noise to propagate. A hardware run would demonstrate that the circuit executes, not that it wins, and at 8 qubits it cannot do more.
\item
  No trainable variational circuit and no quantum convolutional model. This paper covers kernel methods only. The registered convolutional and transfer-learning arms of the clinical protocol and the parameter-matched small CNN of the M-SYNTH protocol were not trained; the recorded reason in both cases is that full fine-tuning across 20 seeds was not feasible on the 8 CPU cores available.
\item
  M-SYNTH is simulated \citep{r18,r19}. It carries correct X-ray transport physics and known ground truth, but no real anatomy, no real detector non-idealities and no real vendor processing. A single lesion size (5.0 mm) and a single lesion density (1.06) were used, and dose was sampled at two levels, not five.
\item
  Qubit count was never varied. Every arm is 8 qubits. The concentration argument \citep{r24} predicts that the picture worsens as n grows, since the 1/2\^{}n floor falls 64-fold from n = 8 to n = 16, but we have not measured it.
\item
  Library versions for the original runs were not recorded, so exact bit-level reproduction of the archived numbers is not currently possible.
\item
  Not assessed: classical assumption diagnostics, calibration curves and decision-curve analysis. None is required for a discrimination comparison and none was run.
\end{enumerate}

\section{Conclusion}

Three active quantum-kernel benchmarks on breast data, all exact classical statevector simulations of 8-qubit fidelity kernels with no quantum component in the inference path, returned no advantage: the quantum kernel was ahead of the cross-validation-selected classical arm in 2 of 200 paired WDBC subsamples and below it on average at every training-set size; on M-SYNTH raw the +0.1736 lead over a tuned RBF-SVM (p = 5.2e-06) fell to +0.0019 (p = 0.937) against a random forest at the same input, with a Laplacian kernel ahead on validation; and the stopping rule failed on both representations once the comparator set was corrected. Two mechanisms by which such a null becomes a positive were measured directly: comparator choice moved the effect by 0.1717 of AUC, and best-of-30 seed selection added 0.0210 on BreastMNIST, each larger than any quantum-classical difference in the active data. One exception is reported because it does not fit: at beta = 2pi the matched closed-form surrogate did not reproduce the M-SYNTH raw kernel (0.5410 against 0.7045), a narrow statement about one approximation family that a random forest's matching accuracy prevents from being read as a quantum effect, and that a learned-surrogate test, left to future work, could still overturn. The withdrawn CMMD arm shows that a dataset-ordering defect can undo an internally registered clinical benchmark. We propose a four-component reporting standard, comprising a validation-selected best classical comparator from a mandatory set, a matched surrogate, concentration diagnostics against the 1/2\^{}n floor, and a selection discount beside every headline, each of which erased an apparent effect in our own measurements.

\subsubsection*{Ethics statement}
This study used only publicly available, de-identified or simulated datasets (WDBC, M-SYNTH, BreastMNIST and CMMD). The authors performed no human-subject research, collected no patient data and required no ethics approval.

\subsubsection*{Data and code availability}
The analysis code, the result files quoted in this article, the frozen split files with their SHA-256 checksums, the environment manifests with library versions, and the registration record (Supplementary S1) are provided to the editors and reviewers during review and will be deposited in a public Zenodo record with a DOI at publication. Lockfiles of the original runs were not archived; the manifests record the environment of the re-analyses. All datasets are public: the Wisconsin Diagnostic Breast Cancer dataset is distributed through the UCI Machine Learning Repository \citep{r17}; M-SYNTH is distributed by the FDA as Regulatory Science Tool RST24AI06.01 under CC0-1.0 \citep{r18}; BreastMNIST is distributed with MedMNIST v2 \citep{r21}; and CMMD is distributed under CC BY 4.0 by The Cancer Imaging Archive \citep{r25}.

\subsubsection*{Declaration of AI use}
LLM-based assistants were used under the authors' supervision to edit and restructure author-written text, to check references and quoted numbers, and to prepare the submission files. All content was reviewed and approved by the authors, who take full responsibility for the manuscript.

\subsubsection*{Conflict of interest}
All authors are employees of Qanary Inc., a company developing mammography screening software. M.Y.E.H. and A.T. are named inventors on a pending US patent application (no. 19/876,140, filed 17 November 2026) concerning input-integrity gating. The authors declare no other competing interests.

\subsubsection*{Author contributions}
R.S.: investigation, software, validation, data curation, writing -- review and editing. A.T.: software, resources, writing -- review and editing. M.Y.E.H.: conceptualisation, methodology, formal analysis, supervision, writing -- original draft, writing -- review and editing.

\subsubsection*{Acknowledgments}
The authors thank the creators and maintainers of the public datasets analysed here. This work was supported by Qanary Inc.~internal funds.

\end{document}